\documentclass[a4paper]{amsart}

\date{March 17, 2016}

\newtheorem{definition}{Definition}
\newtheorem{lemma}{Lemma}
\newtheorem{theorem}{Theorem}
\newtheorem{remark}{Remark}

\def\cz{\mathbb{C}} 
\def\rz{\mathbb{R}} 

\def\gd{\mathfrak{d}}

\def\gx{\mathfrak{x}}
\def\gy{\mathfrak{y}}

\def\rd{\mathrm{d}}
\def\ri{\mathrm{i}}

\def\tr{\mathrm{tr}}

\title[Dipoles in Graphene]{Accumulation Rate of Bound States of Dipoles in
  Graphene}

\author[S. Rademacher]{Simone Rademacher} \email{simone-rademacher@web.de}
\author[H. Siedentop]{Heinz Siedentop} \email{h.s@lmu.de}
\address{Mathematisches Institut\\ Ludwig-Maximilians-Universit\"at
  M\"unchen\\ Theresienstr. 39\\ 80333 M\"unchen\\ Germany}

\begin{document}
\maketitle
\begin{abstract}
  We prove that the bound state energies of the two-dimensional
  massive Dirac operator with dipole type potentials accumulate with
  exponential rate at the band edge. In fact we prove a corresponding
  formula of De Martino et al \cite{Martinoetal2014}.
\end{abstract}
\section{Introduction\label{s1}}

Recently the bound state problem for strained graphene with a dipole
became of interest in the physics literature (De Martino et al
\cite{Martinoetal2014}). This is described effectively by a
two-dimensional massive Dirac operator $D_\phi$ with a dipole
potential $\phi$, i.e.,
\begin{equation}
  \label{eq:dirac}
  D_\phi:=\sigma\cdot p +\sigma_3 - \phi
\end{equation}
with $\sigma:=(\sigma_1, \sigma_2)$ (the first two Pauli matrices),
$p:=(1/\ri)(\partial_1,\partial_2)$, and real valued potential
\begin{equation}
  \label{eq:dipol0}
  \phi := d + s
\end{equation}
where
$$d(x):=
\begin{cases}
  \gd\cdot \tfrac x{|x|} |x|^{-2}& |x|>1\\
  0&|x|\leq1
\end{cases}
$$
with $\gd\in\rz^2$ is the potential of a pure point dipole at the
origin outside the ball of radius one around
the origin. Without loss of generality, we can -- and will from now on
-- pick $\gd:=(b,0)$ with $b>0$, i.e., a multiple of the unit vector
along the first coordinate axis. The potential $s$ will be the --
possibly -- singular part of the potential that is short range in the
sense that $-\Delta -s$ has only finitely many bound states.

It is folklore that the discrete spectrum of $D_\phi$ would be
infinite, if $\phi$ had a non-vanishing Coulomb tail. This is also
true for its three dimensional analogue. However, dimension two and
three differ in the case of a dipole potential: whereas in three
dimensions there are -- for small coupling constant -- only finitely
many eigenvalues in the gap \cite{AbramovKomarov1972},
$\sigma_\mathrm{d}(D_\phi)$ is always infinite in two dimensions. This
has been predicted by De Martino et al \cite{Martinoetal2014} and
proved in \cite{CueninSiedentop2014}. In fact, De Martino et al even
derived a formula for the accumulation rate of the eigenvalues at the
band edge. The purpose of this paper is to prove their formula. To
formulate our result we need some notation:
\begin{definition}
  We write
  \begin{itemize}
  \item \label{eq:N} $N_I(A)$ for the number of eigenvalues
    of a linear operator $A$ in $I\subset\cz$ -- counting
    multiplicity,
  \item $M_b$ for the -- rescaled -- Mathieu operator with
    periodic boundary conditions at $0$ and $2\pi$ defined by
    \begin{equation}
      \label{eq:mathieu}
      (M_bg)(\varphi) = -g''(\varphi)- b\cos(\varphi)g(\varphi),
    \end{equation}
  \item $B_{\ri\nu}$ for the Bessel operator with imaginary
    order defined by
    \begin{equation}
      \label{eq:bessel}
      (B_{\ri\nu}f)(z)= -f''(z)-\frac1zf'(z)-{\nu^2\over z^2}f(z).
    \end{equation}
  \end{itemize} 
\end{definition}
\begin{remark}
  The lowest eigenvalue $a_0$ of the rescaled Mathieu operator $M_b$
  fulfills the transcendental equation (McLachlan
  \cite[3.11.(8)]{McLachlan1947})
  \begin{equation}
    a_0
    = \tfrac{1}{4}\left(- \frac{\tfrac 12 \left(\tfrac b4 \right)^2}{1-\tfrac 14\tfrac{a_0}{4}} - \frac{\tfrac {1}{64} \left(\tfrac b4\right)^2 }{1-\tfrac {1}{16}\tfrac{a_0}{4}} - \frac{\tfrac {1}{576} \left(\tfrac b4\right)^2 }{1-\tfrac {1}{36}\tfrac{a_0}{4}}- ...\right),
  \end{equation}
  in particular $a_0$ is negative for all $b$ (McLachlan
  \cite[3.25. diagram]{McLachlan1947}).
\end{remark}

The above notation allows to formulate our main result.
\begin{theorem} \label{Satz1} Assume that $\phi=d+s$ is real valued
  with $d$ as in \eqref{eq:dipol0}, and that the singular part $s$ is
  relatively compact with respect to $D_0$ and its negative part  $s_-$ even fulfills
  $$\int_{\rz^2}s_-(x)\log(2+|x|)\rd x +\int_0^1s_-^*(\pi t) |\log(t)|] \rd t<\infty$$ 
  and also 
  $$\int_{\rz^2}s^2(x)\log(2+|x|)\rd x +\int_0^1(s^2)^*(\pi t) |\log(t)|] \rd t<\infty.$$ 
  Then
  \begin{equation}
    \label{eq:formel}
    \lim_{E\nearrow1} {N_{(-E,E)}(D_\phi)\over|\log(1-E)|} 
    = \frac1\pi \tr\sqrt{{M_{2b}}_-}.
  \end{equation}
\end{theorem}
Before we embark on the proof we make a few comments:
\begin{description}
\item[Domain] Our condition on the potential $\phi$ assures that $D_\phi$ is
  self-adjoint on $H^1(\rz^2:\cz^2)$ and that the essential spectrum
  of $D_\phi$ is $\rz\setminus(-1,1)$.
\item[Electric Potentials] One possible realization of $\phi$ is to
  think of it as the electric potential of some sufficiently smooth
  and localized charge density $\rho$, i.e.,
  $$\phi(x) = \int_{\rz^3}{\rho(\gy)\rd\gy\over |(x,0)-\gy|}$$
  with vanishing monopole moment, i.e., $\int_{\rz^3}\rho=0$ and to
  assume that the dipole moment $\int_{\rz^3}\gy\rho(\gy)\rd \gy$ of
  $\rho$ points in the direction of the first coordinate axis. (See,
  e.g., Jackson \cite[Chapter 4]{Jackson1965} for a discussion of
  multipole expansions of potentials. Note that we extend the vector
  $x\in\rz^2$ by zero to a vector $\gx=(x,0)\in\rz^3$.)

  Instead of three dimensional densities, we could also allow for
  densities $\rho(\gx)=\rho^{(2)}(x)\delta(x_3)$ that are confined to
  the electron plane.
\item[Point Charges] Our hypothesis excludes point charges located
  directly in the graphene sheet, i.e., the plane in which the
  electrons move. Although it is certainly possible to treat a finite
  number of such singularities with subcritical coupling constants,
  i.e., less than $1/2$, this would require an analysis of the
  compactness properties of $D_\phi^2$ when restricted to functions in
  a ball containing the singularities of $\phi$. We refrain from
  embarking on this subtlety, since it is irrelevant for the long
  range behavior of the potential which determines the asymptotic
  behavior of the eigenvalues.
\end{description}

\section{Proof of the Main Theorem\label{s2}}

Intuitively the long range behavior determines the asymptotic behavior
of the eigenvalues. This motivates to prove the following lemma
including pure dipole potentials.
\begin{lemma}
  For real $a,b\in\rz$, $R\in\rz_+$, $A:=\{x\in\rz^2|\ |x|>1\}$. Set
  $$H_{a,b}:= -\Delta -   {a+b x_1/|x|\over|x|^2}
$$
on $H^2_0(A)$ or $H^2(A)$, i.e., the operator with Dirichlet
respectively Neumann boundary conditions on the boundary of $A$. Then
  \begin{equation}
    \label{hilf}
    \lim_{E\nearrow0}{N_{(-\infty, E)}(H_{a,b})\over|\log(|E|)|} 
    = \frac1{2\pi}\tr\sqrt{(M_b-a)_-}.
  \end{equation}
\end{lemma}

Note that the theorem is only nontrivial, if the sum of the lowest
Mathieu eigenvalue $-\mu$ and $-a$ is negative. This, however, is the
case for $a=0$, since the first Mathieu eigenvalue is always negative
(McLachlan \cite{McLachlan1947}).

We know of two ways proving the lemma. The first relies on
Dirichlet-Neumann bracketing; it adapts an argument that Kirsch and
Simon \cite{KirschSimon1988} developed for direction independent pure
$1/r^2$-potentials to the case of dipole potentials (direction
dependent decay (!)). (See Rademacher \cite{Rademacher2015} for
details.) The method presented here is somewhat different. As we will
see in the proof, the essential ingredient is the short range behavior
of modified Bessel functions with imaginary order which -- we feel --
is more direct. Moreover it is closer to the derivation by De Martino
et al \cite{Martinoetal2014}.

We begin with the proof of the Lemma.
\begin{proof}
  We solve $H_{a,b}\psi = -\lambda \psi$ by separating variables,
  i.e., we make the ansatz $\psi(x)= f(r)g(\varphi)$ in spherical
  coordinates $x_1=r\cos(\varphi)$ and $x_2:=r\sin(\varphi)$ and
  require periodic boundary conditions for $g$, i.e.,
  $g(\varphi)=g(\varphi+2\pi)$. Then $g$ is an eigenfunction of the
  Mathieu operator, i.e., $M_bg= -\mu g$, and 
 $f$ is an eigenfunction
  of the Bessel operator $B_{\ri\sqrt{a+\mu}}f=-\lambda f$ which we
  need to solve with Dirichlet respectively Neumann boundary condition
  at $1$ 
and Dirichlet condition at infinity, depending on whether we
  solve the problem in $H^2_0(A)$ or $H^2(A)$.

  As mentioned above, the Mathieu operator $M_b$ has for all positive
  $b$ at least one negative eigenvalue. Because of the boundary
  condition at infinity we have that the only solutions of
  \eqref{eq:bessel} are
  \begin{equation}
    \label{eq:loesung}
    f(r) = K_{\ri\sqrt{\mu+a}}(\sqrt{\lambda}r).   
  \end{equation}
  (Note that the functions $I_{\ri\sqrt{\mu+a}}(\sqrt{\lambda}\cdot)$
  and $K_{\ri\sqrt{\mu+a}}(\sqrt{\lambda}\cdot)$ are two linearly
  independent solutions of Bessel's equation. 
  However, $I_{\ri\sqrt{\mu+a}}(\sqrt{\lambda}\cdot)$ is
  excluded because of its exponential blow up at infinity (Watson
  \cite[Chapter 7.23, Formula (2)]{Watson1922}).

  Next we note that, since the change from Dirichlet to Neumann
  boundary condition at $1$ is a perturbation of rank one for each
  fixed eigenvalue of the Mathieu equation, we merely need to consider
  the Dirichlet case. Thus we are interested in finding the maximal
  number of nodes that a function
  $K_{\ri\sqrt{\mu+a}}(\sqrt{\lambda}\cdot)$ can have for fixed
  $\mu+a$, assuming that $K_{\ri\sqrt{\mu+a}}(\sqrt{\lambda})=0$ and
  any $-\lambda\leq E$, i.e.,
  \begin{equation}
    \label{eq:n}
    \left|\{r\geq 1|K_{\ri\sqrt{\mu+a}}(\sqrt{\lambda}r)=0, -\lambda\leq E\}\right|.
  \end{equation}
  If $a+\mu\leq0$, the operator $B_{\sqrt{|a+\mu|}}$, has no eigenfunctions and
  the claim is trivially true, i.e., we may assume that $a+\mu>0.$
  Thus, we need to find the maximal $n$ such that
  \begin{equation}
    \label{eq:bed}
    \sqrt{-E}\leq k_{\sqrt{\mu+a},n} 
    = O(\exp(-(n\pi-\phi_{\sqrt{\mu+a}})/\sqrt{\mu+a}))
  \end{equation}
  where $k_{\sqrt{\mu+a},n}$ denotes the zeros of
  $K_{\ri\sqrt{\mu+a}}(\sqrt\lambda\cdot)$ using their asymptotic expansion (see
  \eqref{eq:b1}). Taking the logarithm and dividing by $|\log(-E)|$ yields
  \begin{equation}
    \label{eq:r}
    n/|\log(-E)| \to {\sqrt{\mu+a}\over2\pi}
  \end{equation}
  as $E\nearrow0$.  Since the lower bound can deviate by at most one,
  this is covered as well.
\end{proof}

Next we turn to the proof of the theorem:
\begin{proof}
  We begin by noting that the operator $D_\phi$ is a relatively compact
  perturbation of the free Dirac operator $D_0$. This implies that
  $D_\phi$ has only eigenvalues of finite multiplicity in $(-1,1)$,
  the spectral gap of $D_0$. Furthermore those eigenvalue can only
  accumulate at $-1$ or $1$. Thus, by the spectral theorem
  \begin{equation}
    \label{eq:quadrat}
    N_{(-1,1)}(D_\phi) = N_{(-\infty,0)}(D_\phi^2-1).
  \end{equation}
  which reduces the problem to study the negative eigenvalues of a
  relatively compact perturbation of the Laplacian, since
  \begin{equation}
    \label{eq:quadrat2}
    D_\phi^2 -1 = -\Delta +\phi^2 +(\sigma\cdot p) \phi +\phi(\sigma\cdot p) - 2\sigma_3\phi.
  \end{equation}
  The Schwarz inequality -- followed by the geometric-arithmetic mean
  inequality -- yields for any positive $\epsilon$
  \begin{equation}
    \label{eq:schwarz}
    |2\Re(\psi,\epsilon (\sigma\cdot p)(\epsilon^{-1}\phi)\psi)| \leq \epsilon^2\|p\psi\|^2 +\epsilon^{-2}\|\phi\psi\|^2.
  \end{equation}
  Thus
  \begin{equation}
    \label{eq:oben}
    D_\phi^2 -1 \leq (1+\epsilon^2)p^2 +(1+\epsilon^{-2})\phi^2-2\sigma_3\phi 
  \end{equation}
and
\begin{equation}
  \label{eq:unten}
  D_\phi^2 -1 \geq (1-\epsilon^2)p^2 +(1-\epsilon^{-2})\phi^2-2\sigma_3\phi. 
\end{equation}
Note the lower bound \eqref{eq:unten} is bounded from below for
$\epsilon\in(0,1)$, since both $\phi$ and $\phi^2$ are relative compact
perturbations of $p^2$.

Both right hand sides of \eqref{eq:oben} and \eqref{eq:unten} separate
in two independent one component operators, since $\sigma_3$ is
diagonal. We shall focus on the first component. (As the proof shows
the second component will give the same answer because of the symmetry
of the pure dipole part.)  We write 
\begin{equation}
  \label{eq:pm}
  (1\pm\epsilon^2)h_\pm 
  := (1\pm\epsilon^2)\left(p^2 +\epsilon^{-2}\phi^2-{2\over (1\pm\epsilon^2)}\phi\right)
\end{equation}
for the first components of the right hand side of $\eqref{eq:oben}$
and $\eqref{eq:unten}$. The task is now to estimate
$N_{(-\infty,E)}(h_+)$ from below and $N_{(-\infty,E)}(h_-)$ from
above as $E\nearrow0$. We begin with the lower bound to
$N_{(-\infty,0)}(h_+)$ and write in the spirit of \eqref{eq:sum3}
\begin{align}
  \label{h} {h_\pm} &= -\Delta + V_\pm + W_\pm,\\
  \label{V}
  V_\pm :&= -2(1\pm\epsilon^2)^{-1}d,\\
  \label{W}
  W_\pm :&= \epsilon^{-2}\phi^2-2(1\pm\epsilon^2)^{-1}s.
\end{align}
(Note that the indices $\pm$ at $h$, $V$, and $W$, are just indices
motivated by the signs in \eqref{eq:oben} and \eqref{eq:unten} and not
to the positive part and negative part of an operator as elsewhere in
the paper.)  Thus, by \eqref{eq:sum3}
\begin{equation}
  \label{oben}
  \limsup_{E\nearrow0}{N_{(-\infty,E)}(h_+)\over|\log(|E|)}
  \geq  \limsup_{E\nearrow0}{N_{(-\infty,E)}(-\Delta-(1-\epsilon) V_+)\over|\log(|E|)},
\end{equation}
since $N_{(-\infty,E)}(-\Delta+(1-\epsilon)\epsilon^{-1}W_+)\leq
N_{(-\infty,0)}(-\Delta+(1-\epsilon)\epsilon^{-1}W_+)<\infty$ by
\eqref{eq:shargo1} and thus vanishes when divided by $\log(|E|)$ as
$E\nearrow0$. Finally, we take $\epsilon$ to zero and get
\begin{multline}
  \label{eq:epsilon=0}
  \limsup_{E\nearrow0}{N_{(-\infty,E)}(h_+)\over|\log(|E|)|}\geq  \limsup_{E\nearrow0}{N_{(-\infty,E)}(-\Delta-2d)\over|\log(|E|)|}\\
  \geq
  \limsup_{E\nearrow0}{N_{(-\infty,E)}\left((-\Delta-2d)|_{H_0^2(\{x\in\rz^2||x|>1)\}}\right)\over|\log(|E|)|}
  ={1\over2\pi}\tr\sqrt{{M_{2b}}_-}
\end{multline}
using \eqref{hilf} in the last step. Repeating the argument for the
second component yields the same result. Adding the results for both
components gives the claimed upper bound.

We now turn to the upper bound. We use \eqref{eq:unten} to estimate
the operator from below and thus, the number of eigenvalues from
above. Again the operator decouples into two one-component operators
and we are left with the task to compute twice the number of
eigenvalues of $h_-$ below $-E$ as remarked already above. Next we use
\eqref{eq:sum2} to estimate from above and note -- similarly to the
lower bound -- that the $W_-$ part does not contribute. As above we now get
\begin{multline}
  \label{eq:epsilon=0o}
  \limsup_{E\nearrow0}{N_{(-\infty,E)}(h_-)\over|\log(|E|)|}\leq  \limsup_{E\nearrow0}{N_{(-\infty,E)}(-\Delta-2d)\over|\log(|E|)|}\\
  \leq
  \limsup_{E\nearrow0}{N_{(-\infty,E)}\left((-\Delta-2d)|_{H^2(\{x\in\rz^2||x|>1)\}}\right)\over|\log(|E|)|}
  ={1\over2\pi}\tr\sqrt{{M_{2b}}_-}
\end{multline}
where we estimated by the Neumann operator and used \eqref{hilf}
again.  Doubling the bound because of the two components gives the
desired result.
\end{proof}

\textsc{Acknowledgment:} Thanks go to Sergey Morozov for directing our
attention to \cite{Shargorodsky2013}. We acknowledge partial support
of the Deutsche Forschungsgemeinschaft through its TR-SFB 12
(Symmetrien und Universalit\"at in mesoskopischen Systemen).

\appendix

\section{Some Auxiliary Formulae about the Asymptotic of Bessel
  Functions\label{a1}}

Dunster \cite[Formula 2.8]{Dunster1990} offers the asymptotic formula 
\begin{equation}
  \label{eq:b1}
  k_{\nu,n} = 2 \exp(-(n\pi-\phi_\nu)/\nu)\left(1+{\exp(-2(n\pi-\phi_\nu)/\nu) \over 1+\nu^2} + O(\exp(-4(n\pi-\phi_\nu)/\nu))\right)
\end{equation}
as $n\to\infty$ for the $n-th$ zero (counting from the right) of the
modified Bessel function $K_{\ri\nu}$ for fixed imaginary order. Here
$$ \phi_{\nu}:=\arg\left(\Gamma(1+\ri\nu)\right).$$

\section{Bounds on the number of eigenvalues of Schr\"odinger operators in two dimensions\label{a3}}

Since negative potentials in one and two dimensions generate always at
least one bound state with negative energy, the standard
Lieb-Cwickel-Rosenblum bounds cannot hold. Following earlier works in
the physics literature (see Khuri, Martin, and Wu
\cite{Khurietal2002}), Shargorodsky \cite[Theorem
4.3]{Shargorodsky2013} showed that there is a positive constant $C$
such that for all $V$ the number $N_{-}(V) :=N_{(-\infty,0)}(-\Delta + V)$ of
negative eigenvalues of $-\Delta+V$ is bounded by
\begin{equation}
  \label{eq:shargo1}
  N_-(V)\leq C \left(\int_{\rz^2}V_-(x)\log(2+|x|)\rd x +\int_0^1V_-^*(\pi t) |\log(t)|] \rd t\right) +1
\end{equation}
where the subscript $V_-$ denotes the negative part of $V$ and
$V_-^*$ the spherically symmetric rearrangement of $V_-$.

\section{Known bounds on the number of eigenvalues of sums of
  operators\label{a4}}
   
The number of eigenvalues of the sum of two self-adjoint operators $A$
and$ B$ that are bounded by below and having $\inf
\sigma_\mathrm{ess}(A)=\inf \sigma_\mathrm{ess}(B)=0$ can be estimated
by the sum of the number of eigenvalues of each single operator, i.e.,
for $E<0$
\begin{equation}
  \label{eq:sum1}
  N_{(-\infty,E)}(A+B) \leq  N_{(-\infty,E)}(A) +  N_{(-\infty,E)}(B).
\end{equation}
This formula is a consequence of the minimax theorem and was proved by
Kirsch and Simon \cite[Proposition 4]{KirschSimon1988} resp. Reed and
Simon \cite[p.274, Formula (125)]{ReedSimon1978}.  In particular this
formula applies for Schr\"odinger operators of the form $H= - \Delta
+V +W$ with potentials $V,W$ in $\rz^2$ such that $\inf
\sigma_\mathrm{ess}(H)=0$ \cite[Proposition 5 (i)]{KirschSimon1988}. We
obtain for $\epsilon>0$ and $E<0$
\begin{equation}
  \label{eq:sum2}
  N_{(-\infty,E)}(- \Delta +V +W) \leq  N_{(-\infty,E)}(-\Delta + \tfrac{1}{1-\epsilon}V) +  N_{(-\infty,E)}(-\Delta +\tfrac 1 \epsilon W).
 \end{equation}
 This estimate leads to the lower bound 
 \begin{equation}
   \label{eq:sum3}
   N_{(-\infty,E)}(- \Delta +V +W) \geq  N_{(-\infty,E)}(-\Delta + (1-\epsilon)V) -  N_{(-\infty,E)}(-\Delta -\tfrac {1-\epsilon}{\epsilon} W)
 \end{equation}
 for the number of eigenvalues of $H$ \cite[Proposition 5
 (ii)]{KirschSimon1988}.

\def\cprime{$'$}

\end{document}